%% file: main.tex
\documentclass[sigconf, nonacm]{acmart}
\AtBeginDocument{%
  }

\setcopyright{acmlicensed}
\copyrightyear{2026}
\acmYear{2026}
\acmDOI{XXXXXXX.XXXXXXX}

\acmConference[EASE 2026]{The 30th International Conference on Evaluation and Assessment in Software Engineering}{10–13 June, 2026}{Glasgow, Scotland, United Kingdom}

\acmISBN{978-1-4503-XXXX-X/2018/06}

\usepackage[acronym]{glossaries}
\usepackage{pifont}
\newcommand{\cmark}{\ding{51}} 
\newcommand{\xmark}{\ding{55}} 

\usepackage{tabularx}
\usepackage{booktabs}
\usepackage{multirow,graphicx}
\usepackage{listings}
\usepackage{caption}
\usepackage{xcolor}
\usepackage{float}
\floatstyle{ruled}
\newfloat{listing}{tbp}{lop}

\lstdefinelanguage{Markdown}{
  sensitive=true,
  morecomment=[l]{>},                 
  morecomment=[l]{-},                 
  morecomment=[l]{*},                 
  morestring=[b]`,                    
  morestring=[b]```,                  
  alsoletter={\#\_},
  morekeywords={\#, \#\#, \#\#\#, \#\#\#\#, \#\#\#\#\#},
}

\begin{document}
\newacronym{pp}{pp}{percentage points}
\newacronym{NL}{NL}{Natural Language}
\newacronym{LLM}{LLM}{Large Language Model}
\newacronym{OAS}{OAS}{OpenAPI Specification}
\newacronym{SUT}{SUT}{Service Under Test}
\newacronym{TSL}{TSL}{Test Specification Language}
\newacronym{REST}{REST}{Representational State Transfer}
\newacronym{RBTG}{RBTG}{Requirements-Based Test Generation}
\newacronym{PBMT}{PBMT}{Property-Based Mutation Testing}

\title[RESTestBench]{RESTestBench: A Benchmark for Evaluating the Effectiveness of LLM-Generated
REST API Test Cases from NL Requirements}

\settopmatter{authorsperrow=3}

\author{Leon Kogler}
\email{leon.kogler@casablanca.at}
\orcid{0009-0000-7319-2398}
\affiliation{%
  \institution{CASABLANCA hotelsoftware GmbH}
  \city{Schönwies}
  \country{Austria}
}

\author{Stefan Hangler}
\email{stefan.hangler@casablanca.at}
\orcid{0009-0005-8319-1002}
\affiliation{%
  \institution{CASABLANCA hotelsoftware GmbH}
  \city{Schönwies}
  \country{Austria}
}

\author{Maximilian Ehrhart}
\email{maximilian.ehrhart@casablanca.at}
\orcid{0000-0002-9554-0231}
\affiliation{%
  \institution{CASABLANCA hotelsoftware GmbH}
  \city{Schönwies}
  \country{Austria}
}

\author{Benedikt Dornauer}
\email{benedikt.dornauer@uibk.ac.at}
\orcid{0000-0002-7713-4686}
\affiliation{%
  \institution{University of Innsbruck}
  \city{Innsbruck}
  \country{Austria}
}

\author{Roland Wuersching}
\email{roland.wuersching@tum.de}
\orcid{0000-0002-4979-6180}
\affiliation{%
  \institution{Technical University of Munich}
  \city{Munich}
  \country{Germany}
}

\author{Peter Schrammel}
\email{peter.schrammel@diffblue.com}
\orcid{0000-0002-5713-1381}
\affiliation{%
  \institution{Diffblue Ltd}
  \city{Oxford}
  \country{United Kingdom}
}

\renewcommand{\shortauthors}{Kogler L., Hangler S., Ehrhart M., Dornauer B., Wuersching R., and Schrammel P.}

\begin{abstract}
Existing REST API testing tools are typically evaluated using code coverage and crash-based fault metrics. However, recent LLM-based approaches increasingly generate tests from \gls{NL} requirements to validate functional behaviour, making traditional metrics weak proxies for whether generated tests validate intended behaviour.
To address this gap, we present RESTestBench, a benchmark comprising three REST services paired with manually verified \gls{NL} requirements in both precise and vague variants, enabling controlled and reproducible evaluation of requirement-based test generation. RESTestBench further introduces a requirements-based mutation testing metric that measures the fault-detection effectiveness of a generated test case with respect to a specific requirement, extending the property-based approach of Bartocci et al.~\cite{bartocciPropertyBasedMutationTesting2023}. Using RESTestBench, we evaluate two approaches across multiple state-of-the-art LLMs: (i) non-refinement-based generation, and (ii) refinement-based generation guided by interaction with the running \gls{SUT}. In the refinement experiments, RESTestBench assesses how exposure to the actual implementation, valid or mutated, affects test effectiveness. 
Our results show that test effectiveness drops considerably when the generator interacts with faulty or mutated code, especially for vague requirements, sometimes negating the benefit of refinement and indicating that incorporating actual \gls{SUT} behaviour is unnecessary when requirement detail is high.

\end{abstract}



\received{02 March 2026}
\received[accepted]{21 April 2026}

\maketitle

\section{Introduction}
\textit{\gls{REST}} APIs have long been essential in enterprise systems, enabling scalable integration between services and supporting digital transformation. The adoption of API‑first approaches has accelerated, with a growing number of organizations treating APIs as products that drive business value and foster innovation \cite{mohammedmudassirRoleAPIsModern2024,postmanStateAPIReport}. If faults in deployed API services occur, they may lead not only to service outages but also to data integrity violations and costly production incidents, thereby reinforcing the need for effective and scalable automated API testing.

This need is reflected in the growth of research on \gls{REST} API testing tools between 2017 and 2022 \cite{golmohammadiTestingRESTfulAPIs2024a}. During this period, fuzzing-based approaches became predominant, especially black-box techniques that generate requests from \textit{\gls{OAS}} and evaluate outcomes using oracles such as server errors (\texttt{5xx}) and schema violations \cite{golmohammadiTestingRESTfulAPIs2024a,atlidakisRESTlerStatefulREST2019,arcuriAutomatedBlackWhiteBox2021,martin-lopezRESTestAutomatedBlackbox2021}. Recent advances in \textit{\glspl{LLM}} have shifted the landscape of automated API testing. \glspl{LLM} can interpret \textit{\gls{NL} artifacts} and generate domain-aware assertions, enabling test generators to move beyond purely robustness-oriented checks and towards validating functional correctness and business logic requirements \cite{huangChallengesFuzzingTechniques2024,augustoLargeLanguageModels2025}. 

Emerging tools generate \gls{REST} API tests from \gls{NL} requirements or requirement-like scenarios, either directly from human-defined descriptions (e.g., user stories) or via intermediate \gls{NL} scenarios derived from the \gls{OAS}~\cite{pereiraAPITestGenieAutomatedAPI2024a,zhangLogiAgentAutomatedLogical2025a,panSAINTServicelevelIntegration2025,koglerRESTifAILLMBasedWorkflow2025}, as given in Table~\ref{tab:tool_overview}. Thereby, most tools rely on the \gls{OAS} as the primary source for \gls{LLM}-derived scenarios treated as functional oracles. Unfortunately, since the \gls{OAS} only specifies endpoints, parameters, and response schemas~\cite{andrewsOpenAPISpecification2026} while omitting behavioural semantics and business constraints, it might not reliably serve as a basis for functional requirements.

Apart from that, the shift in using \glspl{LLM} to generate requirements-based tests brings renewed attention  to a long-standing challenge in automated testing: \textit{the oracle problem} \cite{barrOracleProblemSoftware2015}. While current \gls{LLM}-based test generation approaches address the challenge of creating stronger oracles to validate functional requirements, ensuring that tests actually reflect the intended requirements remains an open problem. When tests aim to validate requirements rather than merely detect crashes, traditional metrics such as code coverage or \texttt{5xx} server error counts are insufficient \cite{hossainBriefSurveyOraclebased2022, bartocciPropertyBasedMutationTesting2023}. Consequently, we see that most approaches still rely on manual evaluation (Section \ref{sec:generationApproaches}). Mutation testing offers a stronger adequacy signal than code coverage and crash-based metrics, and has recently been applied to \gls{LLM}-generated \gls{REST} API tests \cite{barradasCombiningTSLLLM2025}. Yet, if mutations are not tied to specific requirements, mutation testing cannot determine whether a test kills mutants for the “right reason”, i.e., because it enforces the intended requirement rather than incidental implementation behaviour. Bartocci et al.~\cite{bartocciPropertyBasedMutationTesting2023} introduce \textit{\gls{PBMT}}, emphasizing that mutants are only meaningful when tied to a specific property: a mutant is considered killed in a meaningful way only if its execution triggers a violation of that property. Existing collections of services used in \gls{REST} API testing research (including the Public \gls{REST} API Benchmark by Decrop et al. \cite{decropPublicBenchmarkREST2025}) document which APIs have been used for evaluation, but do not provide verified \gls{NL} requirements or requirement related mutations that would allow controlled measurement of test generation from requirements. Consequently, human judgment remains necessary to assess whether generated tests meaningfully validate intended requirements, limiting reproducibility, and comparability across tools.

\input{tables/tool_overview} 

To address these limitations, we introduce \textit{RESTestBench}, a benchmark that explicitly separates requirement engineering from test generation and assesses test adequacy using specific mutations relevant to the requirement, drawing inspiration from Bartocci et al. and their \gls{PBMT} definition. By grounding evaluation in validated requirements, the effectiveness of test generation approaches can be assessed quantitatively without conflating ambiguity in requirement generation with errors in test generation. Assessing the adequacy of generated tests by using \gls{PBMT} is not only a more challenging metric than normal mutation testing \cite{bartocciPropertyBasedMutationTesting2023}, it also allows us to evaluate if requirement based testing approaches that interact with the \textit{\gls{SUT}}, are able to distinguish between a valid implementation and an already mutated implementation. While regression-based tools assume the correctness of the current \gls{SUT} implementation, requirements-based approaches should distinguish invalid from valid behaviour, as their oracle is grounded in the requirements rather than in the implementation. Therefore, our benchmark enables the evaluation of test generation approaches that encounter the actual behaviour of the \gls{SUT} in two settings: one in which tests are generated based on the valid implementation, and one in which tests are generated based on a mutated implementation. This design enables systematic and reproducible comparison of requirements-based generation strategies without relying on human judgement as the primary oracle. 

Overall, this work makes the following contributions:
\begin{itemize}
    \item\textbf{RESTestBench}: A requirements-based REST API test generation benchmark comprising three REST services, 106 human-validated \gls{NL}-requirements at two different levels of detail, 228 manually designed requirements-based mutations, and a framework that enables straightforward integration and evaluation of new generation approaches.
    \item\textbf{Two experiments conducted using RESTestBench}:
    \begin{enumerate}
        \item An evaluation of a simple single-step approach serving as a baseline for non-refinement-based tools.
        \item An evaluation of a refinement-based approach that incorporates knowledge of the actual behaviour of the \gls{SUT}. Specifically, we measure whether actual behaviour influences test effectiveness, as observed in prior work on \gls{LLM}-generated test oracles \cite{konstantinouLLMsGenerateTest2024}.
    \end{enumerate}
\end{itemize}

\section{Background and Related Work}

\subsection{LLM-Based REST API Test Generation Approaches}
\label{sec:generationApproaches}
Over the past decade, the number of automated approaches for testing \gls{REST} APIs has increased substantially~\cite{golmohammadiTestingRESTfulAPIs2024a}. Until roughly 2022, most practical automated \gls{REST} API testing approaches were variants of \emph{property-based} and \emph{model-based} testing. Representative tools include EvoMaster~\cite{arcuriAutomatedBlackWhiteBox2021}, RESTest~\cite{martin-lopezRESTestAutomatedBlackbox2021}, and RESTler~\cite{atlidakisRESTlerStatefulREST2019}.

Despite the prevalence of property- and model-based testing, automatically deriving test oracles from textual documentation (e.g., API descriptions or requirements in natural language) remained a significant challenge. In general software testing, the oracle problem has long been recognized as difficult when relying on unstructured text alone, since human interpretation is required to map documentation to expected behaviour~\cite{barrOracleProblemSoftware2015} and existing \gls{REST} API tools predominantly generate tests from machine-readable specifications rather than textual requirements~\cite{alonsoAGORAAutomatedGeneration2023}. Meanwhile, in broader software engineering, \textit{\gls{RBTG}} has been shown to align testing closely with user intent and quality assurance goals by constructing test cases directly from \gls{NL} requirements~\cite{yangRequirementsBasedTestGeneration2025}.
Recent advances in \glspl{LLM} have significantly lowered the barrier to leveraging \gls{NL} artifacts for \gls{REST} API testing. In the following, we examine current \gls{LLM}-based \gls{REST} API testing tools that derive, interpret, and/or translate \gls{NL} requirements into executable test cases; these approaches are summarized in Table~\ref{tab:tool_overview}.\\

\noindent\textbf{APITestGenie} \cite{pereiraAPITestGenieAutomatedAPI2024a} combines an \gls{OAS} with human-written \gls{NL} requirements and uses single-shot \gls{LLM} prompting to generate executable \textit{JUnit} tests, treating the provided requirements as the oracle. The experimental evaluation focuses on executability and perceived usefulness, with manual validation recommended.

\noindent\textbf{LogiAgent} \cite{zhangLogiAgentAutomatedLogical2025a} adopts a multi-agent workflow that derives \gls{NL} scenarios from the \gls{OAS}, executes them stepwise, and validates responses using an \gls{LLM} as a judge. Effectiveness is measured via code coverage and detected failures, while semantic correctness relies on manual annotation.

\noindent\textbf{RestTSLLM} \cite{barradasCombiningTSLLLM2025} employs a two-stage \gls{LLM} pipeline: first generating a \textit{\gls{TSL}}-based test specification from the \gls{OAS}, then translating it into executable tests. Evaluation includes execution success, coverage, and mutation score, but does not explicitly assess alignment with intended requirements. 

\noindent\textbf{SAINT} \cite{panSAINTServicelevelIntegration2025} is a white-box approach that infers endpoints and dependencies from source code and refines them into tests using \gls{LLM} agents. Oracles are embedded implicitly in generated scenarios, and evaluation combines automated metrics with developer feedback.

\noindent\textbf{RESTifAI} \cite{koglerRESTifAILLMBasedWorkflow2025} generates happy-path and negative test scenarios from an \gls{OAS} using \glspl{LLM}, focusing on robustness. Effectiveness is assessed through coverage and detected server errors, with additional expert-based validation.

Across the reviewed approaches, two fundamental generation strategies can be distinguished. Non-refinement approaches (e.g., APITestGenie, RestTSLLM) generate test cases in a single-step manner from the \gls{OAS} and/or \gls{NL} requirements without interacting with the \gls{SUT} during generation. In contrast, refinement-based approaches (e.g., LogiAgent, SAINT, RESTifAI) incorporate feedback loops in which intermediate requests are executed against the \gls{SUT}, and the observed responses are used to iteratively refine scenarios, parameters, or assertions. These strategies therefore differ primarily in whether test generation is performed statically from the specification or dynamically informed by the runtime behaviour of the service. The intention of the experiments in this paper is not to report comprehensive comparison results for existing tools, but to show how the benchmark can be used to compare fundamentally different generation approaches in a controlled way under uniform experimental conditions. 

\subsection{Metrics and their Problem for Requirements-Based Testing} 
In 2023, Golmohammadi et al.~\cite{golmohammadiTestingRESTfulAPIs2024} conducted a systematic literature review (RQ4) to identify the evaluation metrics commonly used by predominantly non-LLM-based \gls{REST} API testing techniques. Based on that, they group these metrics into three main categories: coverage metrics (e.g., schema-based and code coverage), fault-detection metrics (e.g., HTTP 5xx errors), and performance metrics (e.g., response time and latency). Despite the recent advances in \gls{LLM}-based \gls{REST} API test generation (Section~\ref{sec:generationApproaches}) have not influenced the choice of evaluation strategies, the first two categories are still predominant, as identifiable in Table~\ref{tab:tool_overview}. 

\textit{Coverage metrics} are reported by all approaches discussed in Section~\ref{sec:generationApproaches}, except APITestGenie. Such metrics quantify how extensively the system is exercised, not whether the generated tests actually check the intended functionality~\cite{vanoverbergheStateCoverageSoftware2012}. Furthermore, the correlation between test suite effectiveness and coverage is not strongly given, as quantified by Inozemtseva and Holmes \cite{inozemtsevaCoverageNotStrongly2014}.

\textit{Fault-detection metrics.} Li and Offutt~\cite{liTestOracleStrategies2017} define an adequate test case as one that not only triggers and propagates faulty behaviour but also includes a test oracle capable of revealing the fault. Most of the tools discussed in Section~\ref{sec:generationApproaches} aim to generate increasingly sophisticated oracles to detect a broader range of faults. In practice, however, many of these approaches still rely on server-side error counts (e.g., HTTP 5xx responses) to evaluate test effectiveness. Such metrics are insufficient for assessing the true effectiveness of test oracles when the goal is to validate the functional behaviour of the \gls{SUT}.

Only RestTSLLM~\cite{barradasCombiningTSLLLM2025} integrates a mutation score, where small, faults known as \textit{mutants} are systematically introduced directly into the system to evaluate whether the test suite, particularly API tests, can detect the resulting behavioural deviations. A higher mutation score therefore serves as an indicator of test effectiveness and provides a stronger link to the ability of the tests to detect meaningful defects in \gls{REST} API implementations \cite{sanchezMutationTestingPractice2024, MutationTestingAdvances2019, papadakisAreMutationScores2018a}. Thereby, classical mutation evaluation primarily asks: “If a developer introduced a small coding mistake, would our tests detect it?”. It does not determine whether the tests genuinely enforce the intended system behaviour as specified by requirements, but rather whether they can distinguish the current implementation from syntactically perturbed variants, as is typically done in regression testing. Therefore, PBMT proposed by Bartocci et al.~\cite{bartocciPropertyBasedMutationTesting2023} aims to bridge this gap by evaluating test suites with respect to \textit{property-based} mutations. In our context, these properties correspond directly to formalized requirements. Compared to the established metrics in REST API testing, \gls{PBMT} offers the advantage of assessing whether tests validate the intended system requirements rather than merely detecting arbitrary output differences. By treating requirements as explicit constraints during mutant evaluation, PBMT further prevents mutation scores from being inflated by mutants that are irrelevant to the requirement under consideration. Overall, PBMT represents a promising and more requirement-aligned metric that deserves further attention in this and future work.

\section{RESTestBench Conception} 
\label{sec:benchmark_conception}

\begin{figure}[h]
    \centering
    \includegraphics[width=\columnwidth]{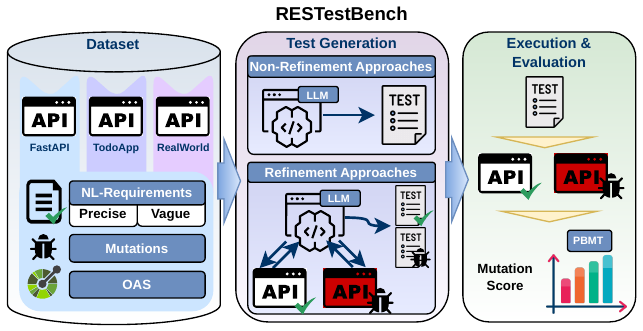}
    \caption{RESTestBench overview}
    \label{fig:benchmark-overview}
\end{figure}

RESTestBench is designed to evaluate the effectiveness of \gls{LLM}-based black-box approaches in translating a functional \gls{NL} requirement, supported by an \gls{OAS}, into a single executable test case. To this end, the benchmark comprises a curated set of \gls{REST} API services (Section~\ref{sec:service_selection}) and a manually defined collection of functional \gls{NL} requirements (Section~\ref{sec:nl_req}). The fault-detection effectiveness of the generated requirements-based test cases is assessed using \gls{PBMT}~\cite{bartocciPropertyBasedMutationTesting2023} with manually defined and validated mutations (Section~\ref{sec:mutation_construction}). Figure~\ref{fig:benchmark-overview} illustrates the overall benchmark workflow, including the generation of test cases from the requirements and the \gls{OAS}, their execution against the \gls{SUT}, and the subsequent evaluation of test outcomes using mutation scores. The benchmark is designed to grow over time: additional services, requirements and mutations can easily be integrated into the current benchmark framework.

The repository and guidelines can be accessed here: 
\begin{center}
    \textit{\url{https://github.com/casablancahotelsoftware/RESTestBench}}.
\end{center}

\subsection{Selection of suitable REST API Services} \label{sec:service_selection}

\input{tables/list_of_services}

The benchmark services were selected to support controlled, reproducible experimentation while remaining representative of real-world \gls{REST} backends. Practical relevance refers to selecting services that resemble production-style \gls{REST} backends in terms of API surface, security mechanisms, and that include complex operation dependencies requiring stateful request sequences.

At a minimum, candidate services must be open-source and provide an \gls{OAS}. Moreover, we require sufficient project maturity for realistic engineering practices and stable behaviour. As a transparent proxy we report GitHub stars and forks, given in Table ~\ref{tab:services}.

Beyond these minimum constraints, we prioritised services whose domain logic is non-trivial and whose endpoints exhibit meaningful dependencies (e.g., authentication prerequisites, ownership constraints, and state-dependent workflows), because these properties enable requirements that go beyond endpoint reachability and require semantically rich assertions.

A further prerequisite is that the \gls{OAS} aligns with the actual service implementation so that tools can derive valid requests and create correct test cases. For services where the provided \gls{OAS} deviated from the observed runtime behaviour (e.g., mismatched status codes, missing/incorrect fields, or parameter constraints), we performed targeted manual corrections to obtain a specification that matches the evaluated service implementation, ensuring that all tools can in principle generate correct tests. Future work could investigate how the quality of the \gls{OAS} influences test quality.

Applying these constraints, we selected the following three services as \textit{\glspl{SUT}}:\\

\noindent \textbf{FastAPI full-stack template}~\cite{FastapiFullstackfastapitemplate2026} is a production-oriented, Python-based FastAPI template service with a PostgreSQL-backed data model, JWT-based security components, and multi-entity workflows (e.g., hierarchical resource ownership and administrative/superuser operations). We selected it because it is widely adopted in the FastAPI ecosystem and provides a realistic, dependency-rich baseline for business-logic requirements.\\

\noindent \textbf{TodoApp}~\cite{fowlerDavidfowlTodoApp2026} is a compact .NET Minimal API service with authentication and domain-specific dependencies that yields a controlled lower-complexity baseline. We selected it because it enables direct comparison between simpler CRUD-centric behaviour and more complex multi-entity workflows and was previously used as an evaluation subject in Barradas et al.~\cite{barradasCombiningTSLLLM2025}.\\

\noindent \textbf{NestJS RealWorld app}~\cite{jakobLujakobNestjsrealworldexampleapp2026} is a NestJS backend (TypeScript) that implements the RealWorld API specification \cite{RealworldappsRealworldMother}, including articles, comments, profiles, favourites, and follower relationships, making it well-suited for complex requirements-based testing.\\

\subsection{Natural Language Requirements} \label{sec:nl_req}

In our benchmark, an \gls{NL} requirement is an unstructured textual statement in human language that specifies the expected externally observable functional behaviour of a \gls{REST} API. The use of natural language reflects common industrial practice, where requirements are frequently documented informally and exhibit varying degrees of precision and completeness, often leading to ambiguity and underspecification \cite{vanlamsweerdeGoalorientedRequirementsEngineering2000,zowghiErratumInterplayConsistency2004}. As the benchmarked services do not provide complete requirements documentation, requirements are reverse-engineered from existing integration tests, where available, or derived through manual inspection of the service source code and its business logic, ensuring that each requirement reflects the actual behaviour of the evaluated system. To validate this alignment, we manually created a golden test for each requirement, verifying that it passes on the original implementation. Each requirement is structured so that, in combination with the \gls{OAS}, it can be transformed into exactly one \gls{REST} API test case.

To study how requirement detail affects test generation effectiveness, we provide two variants of each requirement: one precise and one vague. This distinction is grounded in established requirements engineering literature differentiating goal-level specifications from operational or scenario-based descriptions \cite{vanlamsweerdeGoalorientedRequirementsEngineering2000,jacobsonObjectorientedSoftwareEngineering2005}.

The \textbf{vague} variant is a high-level informal statement that captures only the essential intent of the requirement, providing the minimal information a human tester would need to produce a test case with a still clear and verifiable objective. It omits any preconditions or setup needed for validation, specifying only the testing objective and the expected outcome at a conceptual level. This formulation corresponds to goal-oriented or early-phase requirements that intentionally abstract from operational detail and may remain partially incomplete \cite{vanlamsweerdeGoalorientedRequirementsEngineering2000,zowghiErratumInterplayConsistency2004}. Such underspecification reflects realistic industrial requirements and poses known challenges for automated test generation from \gls{NL} descriptions \cite{aroraGeneratingTestScenarios2024}.

The \textbf{precise} variant, in contrast, specifies all preconditions required to validate the intended test objective. It details the exact sequence of operations, their dependencies, and the parameters relevant for subsequent requests. Additionally, it defines the expected results concretely, including response fields and values that must be verified. Compared to the vague variant, this representation functions more like a scenario-based description of a test case. Its structure is more aligned with use cases and scenario-based specifications \cite{cockburnWritingEffectiveUse2012, jacobsonObjectorientedSoftwareEngineering2005}. By specifying all required preconditions, interaction steps, and expected outputs, the precise variant fulfills established quality characteristics of testable requirements, including completeness, unambiguity, and verifiability \cite{wiegersSoftwareRequirements2013}. Listing~\ref{lst:requirement} shows a concrete example of both variants.

\begin{listing}[!t]
\caption{JSON Requirement Example from FastAPI}
\label{lst:requirement}
\begin{lstlisting}[
    basicstyle=\ttfamily\scriptsize,
    breaklines=true,
    breakatwhitespace=false,
    keepspaces=true,
    columns=fullflexible,
    numbers=left,
    numberstyle=\tiny,
    numbersep=4pt,
    xleftmargin=16pt
]
{
  "id": 28,
  "service": "fastapi",
  "requirement_vague": "A superuser (admin@example.com / password123) creates 2 random users and retrieves all. Verify the response.",
  "requirement_precise": "Authenticate as the superuser (use admin@example.com / password123) to obtain an access token. Using that superuser access token, create two distinct users via the public create-user API (each with a unique email and password). Store the returned user IDs and emails. Then, using the same superuser access token, retrieve the list of all users. Verify the response status is 200. Verify the response JSON contains a top-level 'data' collection and a top-level 'count' key. Verify that both created users appear in the 'data' collection by checking that their emails are present. Verify that for each created user, the email in the list matches the email used during creation.",
  "mutants": [
    ...
  ]
}
\end{lstlisting}
\end{listing}

RESTestBench covers a wide range of requirement complexity to support differentiated evaluation. Scenarios range from single-operation requirements to multi-step sequences with intricate dependencies among operations and parameters. Verification tasks vary from simple status code checks to detailed assertions of nested response fields that depend on earlier operations. The benchmark also includes challenging functional contexts, such as complex CRUD workflows, authentication and authorization procedures, and cascading state changes. By addressing both operational and verification difficulty, RESTestBench ensures that test generation methods are assessed across realistic and representative REST API behaviours.

\subsection{Manual Mutation Construction}  \label{sec:mutation_construction}

To assess the effectiveness of requirements-based test generation, we employ manually defined and validated mutations, following the principles of Property-Based Mutation Testing~\cite{bartocciPropertyBasedMutationTesting2023}. Each requirement is formalized as a property, and for each property we define a set of mutations that lead to violations of that property.

Compared to automated mutation seeding, manual mutation design ensures the following:
\begin{enumerate}
    \item \textit{Propagation}: According to the RIP model~\cite{ammannIntroductionSoftwareTesting2017}, a fault must propagate to an observable output to be revealed~\cite{liTestOracleStrategies2017} by an oracle. By explicitly defining and executing mutations, we ensure that violations are observable through the \gls{REST} API.
    
    \item \textit{Relevance}: Mutations are constructed to directly impact the satisfaction of the tested property~\cite{bartocciPropertyBasedMutationTesting2023}. Existing automated mutation tools are, to our knowledge, unable to generate mutants that meaningfully affect \gls{NL} requirements.
    
    \item \textit{Non-equivalency~\cite{demeyerFormalVerificationDeveloper2020}}: Since identifying equivalent mutants is undecidable, mutations are manually defined to ensure behavioural divergence from the original system for at least one input.
    
    \item \textit{Non-subsumption~\cite{MutationTestingAdvances2019}}: Mutations are designed to avoid subsumption, where non-equivalent mutants produce identical propagated outputs and thus provide no additional evaluation power.
    
    \item \textit{Realism}: Common mutation operators (e.g., arithmetic, logical, or syntactic changes~\cite{colesPITPracticalMutation2016, groceExtensibleRegularexpressionbasedTool2018}) are largely semantics-agnostic. In contrast, our manually defined mutants (Listings~\ref{lst:mut_email_swap}) capture realistic faults that static mutation tools are unlikely to produce.
\end{enumerate}

\begin{listing}[!t]
\caption{Valid implementation of the get\_users operation of the FastAPI service}
\label{lst:valid_implementation}
\begin{lstlisting}[
  language=Python,
  basicstyle=\ttfamily\scriptsize,
  breaklines=true,
  breakatwhitespace=false,
  keepspaces=true,
  columns=fullflexible,
  numbers=left,
  numberstyle=\tiny,
  numbersep=4pt,
  xleftmargin=16pt
]
@router.get(dependencies=[Depends(get_current_active_superuser)], response_model=UsersPublic)
def read_users(session: SessionDep, skip: int = 0, limit: int = 100) -> Any:
    count_statement = select(func.count()).select_from(User)
    count = session.exec(count_statement).one()
    statement = select(User).offset(skip).limit(limit)
    users = session.exec(statement).all()
    return UsersPublic(data=users, count=count)
\end{lstlisting}
\end{listing}

Listing~\ref{lst:valid_implementation} presents an example endpoint definition from the full-stack-fastapi-template service used in our benchmark. Listing~\ref{lst:requirement} specifies the corresponding requirement the endpoint in Listing~\ref{lst:valid_implementation} is expected to fulfill and that the generated test case must validate. Listing~\ref{lst:mut_email_swap} illustrates a manually designed mutation that conforms to the above defined criteria. In addition, and in accordance with these criteria, we deliberately restrict the mutation set to a minimal yet representative subset that targets the most critical test assertions associated with the requirement. This design choice ensures a manageable benchmark runtime, given the substantial computational cost of test generation and execution.

\begin{listing}[!t]
\caption{Example mutation of returned user data ensuring the test not only checks the presence of emails in the response, but also if each email is correctly assigned to the corresponding user}
\label{lst:mut_email_swap}
\begin{lstlisting}[
  basicstyle=\ttfamily\scriptsize,
  breaklines=true,
  breakatwhitespace=false,
  keepspaces=true,
  columns=fullflexible,
  numbers=left,
  numberstyle=\tiny,
  numbersep=4pt,
  xleftmargin=16pt
]
- return UsersPublic(data=users, count=count)
+ emails = [u.email for u in users]
+ swapped = [
+   {**u.__dict__, 'email': emails[(i + 1) % len(emails)]}
+   for i, u in enumerate(users)
+ ]
+ return UsersPublic(data=swapped, count=count)
\end{lstlisting}
\end{listing}

\subsection{Evaluation Process} \label{sec:evaluation_setup}
We adopt the \gls{PBMT}~\cite{bartocciPropertyBasedMutationTesting2023} definition of mutation score (Definition~III.4) and apply it under a restriction to singleton test sets. In our setting each requirement \(\phi_i\) is evaluated using a test suite \(T_i = \{t_i\}\) containing a single test case. A mutant is considered \(\phi_i\)-killed according to Definition~III.1 if it is killed by the corresponding test \(t_i\). We define \(M_{\phi_i}\) as the set of mutations defined for \(\phi_i\), and let
\(k_{\phi_i}\) denote the number of mutations \(\phi_i\)-killed by the corresponding test \(t_i\). Let \(M\) denote the set of all mutations in the benchmark and \(k_\phi\) denote the number of all \(\phi_i\)-killed mutations in total. The overall requirements-based mutation score \(MS_\phi\) is defined as:
\begin{equation}
MS_\phi =
\frac{k_\phi}
     {\left|M\right|}
\label{eq:mutation-score}
\end{equation}
To compute the mutation score for refinement-based approaches where the generated test case may depend on the concrete implementation of the \gls{SUT}, we define \( t_i^{\text{valid}} \) as the test case generated from requirement \( \phi_i \) using the valid implementation of the \gls{SUT}. Furthermore, for each mutant \( m_j \in M_{\phi_i} \), we define \( t_i^{m_j} \) as the test case generated based on the \( j \)-th mutated \gls{SUT} of requirement \( \phi_i \).

For \emph{non-refinement-based} approaches, test generation is independent of the actual implementation.
Thus, \( t_i^{\text{valid}} = t_i^{m_j}\), and we evaluate the \(k_{\phi_i}\) based on \( t_i^{\text{valid}} \) and calculate total mutation score as defined in Equation~\ref{eq:mutation-score}. 

\begin{figure}[h]
    \centering
    \includegraphics[width=\columnwidth]{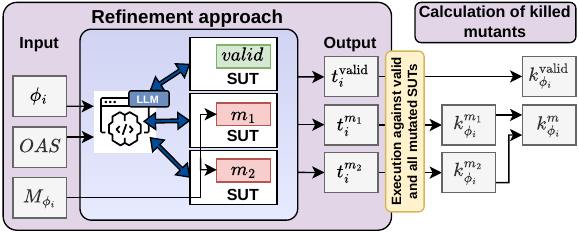}
    \caption{Visualization of the evaluation of \(k_{\phi_i}^{\text{valid}}\) and \(k_{\phi_i}^m\) for a single requirement \(\phi_i\) in refinement-based test generation approaches.}
    \label{fig:refinement_evaluation}
\end{figure}

For \emph{refinement-based} approaches, test generation may depend on the implementation, so we assume that \(t_i^{\text{valid}} \neq t_i^{m_j}\). To compare fault-detection effectiveness fairly between the single \(t_i^{\text{valid}}\) and possible multiple \(t_i^{m_j}\), we compute the average number of mutants killed by all \(t_i^{m_j}\). Specifically, let \(k_{\phi_i}^{m_j}\) denote the number of mutations \(\phi_i\)-killed by \(t_i^{m_j}\), and define

\[
k_{\phi_i}^m = \frac{1}{|M_{\phi_i}|} \cdot \sum_{m_j \in M_{\phi_i}} k_{\phi_i}^{m_j}
\]

i.e., the average number of \(\phi_i\)-killed mutations across all mutant-specific tests. Similarly, let \(k_{\phi_i}^{\text{valid}}\) denote the number of mutations killed by \(t_i^{\text{valid}}\). Figure~\ref{fig:refinement_evaluation} demonstrates the differences between \(k_{\phi_i}^{\text{valid}}\) and \(k_{\phi_i}^m\) for a single requirement.

Summing over all requirements, we define

\[
k_\phi^m = \sum_i k_{\phi_i}^m, \quad
k_\phi^{\text{valid}} = \sum_i k_{\phi_i}^{\text{valid}}.
\]

Based on this distinction, we compute two mutation score values for refinement-based approaches:
\[
MS_\phi^\text{valid} = \frac{k_{\phi}^{\text{valid}}}{|M|} \quad\text{and}\quad
MS_\phi^\text{m}   = \frac{k_{\phi}^m}{|M|}
\]

By computing both scores for refinement-based approaches, RESTestBench extends standard mutation testing by explicitly analyzing the effect of the actual implementation on test effectiveness. In particular, this distinction allows us to investigate whether a tool that incorporates the concrete behaviour of the \gls{SUT} can correctly differentiate between valid and mutated implementations for a given requirement, and how differences between the valid and mutated implementations influence the ability of the generated test cases to detect faults.
\input{tables/overview_nr_req_mutations}
Table~\ref{tab:overview_requirements} summarizes, for each service, the number of requirements and the total number of associated mutations defined in the benchmark. 

To enable standardized and comparable evaluation of different approaches, we define a minimal Python interface consisting of the functions \textit{\(generate\_test\)} and \textit{\(execute\_test\)}, which must be implemented for each new approach. This abstraction ensures independence from specific test languages and frameworks. To guarantee independent test generations (only relevant for refinement-based generation approaches) and test executions, the \gls{SUT} is reset to a well-defined initial state prior to each generation process and each test execution. To track the cost of test generation the \textit{\(generate\_test\)} function is required to return the generation cost in USD to the caller. To validate the correctness of mutation insertion and benchmark execution, we developed a set of golden tests that achieve a full mutation score on the benchmark.

All results are made publicly available here 
\begin{center}
\textit{\url{https://github.com/casablancahotelsoftware/RESTestBench/tree/master/results}}.
\end{center}

\section{Experiments}
\label{sec:experiments}

\begin{figure*}[!t]
    \centering
    \includegraphics[width=0.80\linewidth]{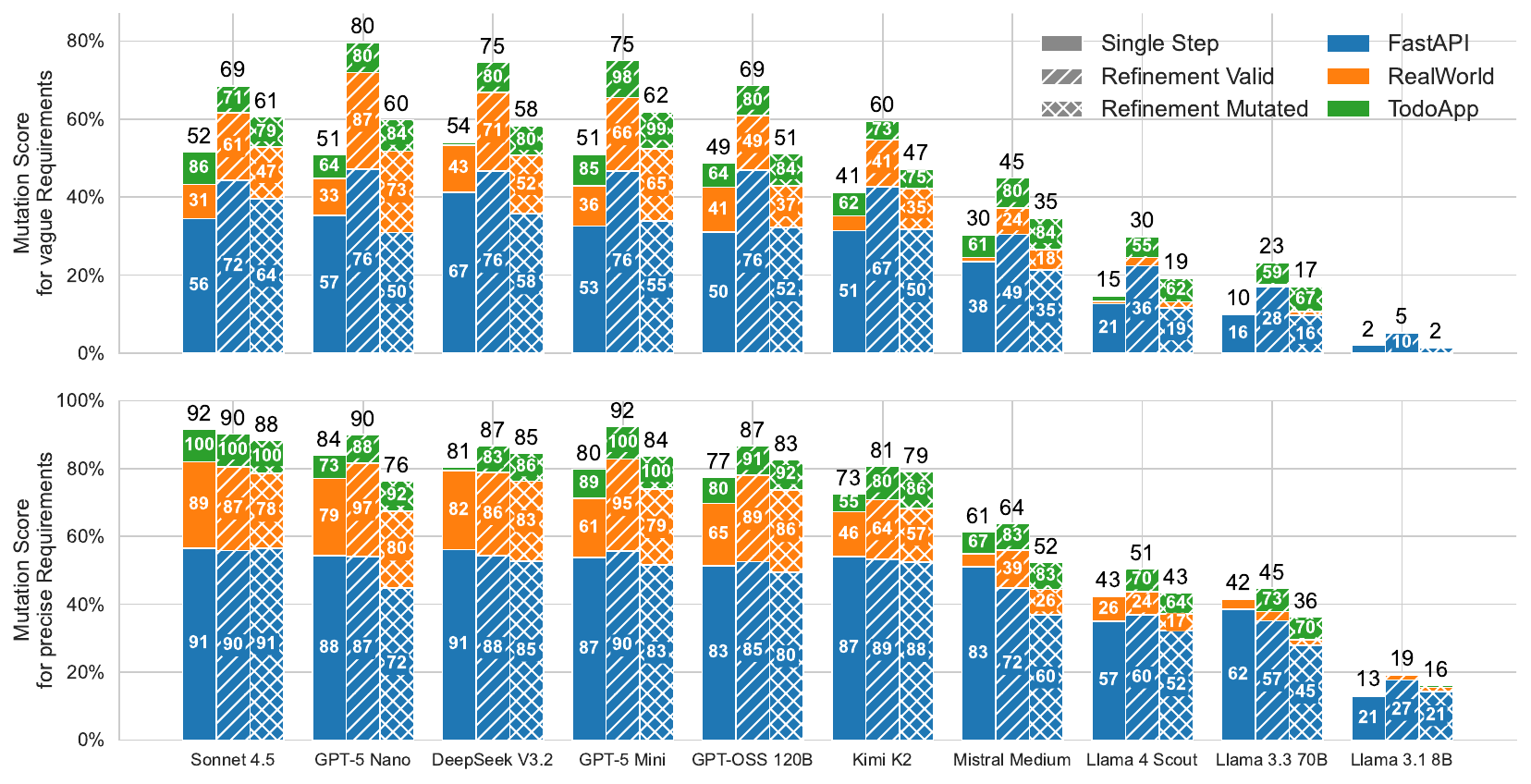}
    \caption{Mutation scores for \emph{vague} and \emph{precise} requirements comparing single-step \(MS_\phi\) (solid bars) with refinement-based scores \(MS_\phi^{\text{valid}}\) (forward-hatched) and \(MS_\phi^{\text{m}}\) (cross-hatched). Bars are stacked by service (FastAPI, RealWorld, TodoApp); segment height reflects weighted service contribution and segment percentages are per-service mutation scores.}
    \label{fig:zs_vs_refinement_vague}
\end{figure*}

This section presents two controlled experiments evaluating \gls{LLM}-based \gls{REST} API test generation on RESTestBench. As discussed in Section~\ref{sec:generationApproaches}, existing approaches can broadly be classified as either \emph{non-refinement} or \emph{refinement-based}. We therefore evaluate one representative strategy for each class under controlled conditions to establish a baseline for subsequent comparisons.

To ensure a fair comparison, both experiments use the same core input artifacts (the \gls{NL} requirement, the \gls{OAS}, and the service base URL), the same target output format (\textit{pytest} test cases), and the same evaluation procedure defined in Section~\ref{sec:evaluation_setup}. Across both experiments, we measure the mutation score achieved on RESTestBench as well as the cost of test generation. In addition, both \emph{precise} and \emph{vague} requirement variants are evaluated to assess how requirement granularity influences test effectiveness.

\input{tables/list_of_models}
Table~\ref{tab:llm_models} lists the ten state-of-the-art \glspl{LLM} evaluated in both experiments. Models were selected along the Pareto frontier of cost versus quality on the Azure AI Foundry model benchmark, ranging from cost-efficient models (e.g., Llama~3.1~8B) to frontier models (e.g., Sonnet~4.5), and spanning six vendors (Anthropic, DeepSeek, Meta, Mistral AI, Moonshot AI, and OpenAI) with six open-source (OS) and four closed-source (CS) models.\footnote{We used \textit{Llama 3.1 8B} instead of the newer \textit{Llama 3.3 8B} because the latter was not deployable in Azure AI Foundry at the time of our experiments.} Due to the substantial time and computational cost of the full benchmark, each experiment was executed three times; we report averaged results with a maximum observed deviation of $3.4$~\gls{pp} in overall mutation score across runs.

\subsection{Experiment 1: Single-Step Generation}
\label{sec:exp1}

The first experiment evaluates a non-refinement, single-step strategy. In this setting, the model receives a zero-shot prompt and generates the final test in one inference call, without interacting with the running \gls{SUT} during generation. The prompt (Listing~\ref{lst:zero_shot_gen_prompt}) takes three inputs: the \gls{NL} requirement description, the complete \gls{OAS}, and the service base URL. The generator neither executes the test nor contacts the \gls{SUT}. Consequently, the produced test is independent of the actual runtime behaviour of the service and is not influenced by potential faults in the current implementation.

\paragraph{Results.}
The single-step mutation scores for all models are shown in Figure~\ref{fig:zs_vs_refinement_vague}, using the solid bars (\(MS_\phi\)). For precise requirements, scores span a wide range, from 13\% (Llama 3.1 8B) to 92\% (Sonnet 4.5), with three models exceeding 80\% (Sonnet 4.5, GPT-5 Nano, and DeepSeek V3.2). For vague requirements, scores drop substantially for every model, typically by 26 to 40 percentage points, and the strongest models converge to a narrower range of 49\%--54\%. Sonnet 4.5 performs very strongly on precisely defined requirements, while Llama 3.1 8B nearly fails on vague requirements, reaching only 2\%.

Overall, these results show that requirement granularity strongly influences single-step generation. More detailed requirements substantially improve the effectiveness of all evaluated models, whereas underspecified requirements remain challenging even for frontier models.

\begin{listing}[!t]
\caption{Shortened zero-shot generation prompt template. Placeholders are filled per requirement.}
\label{lst:zero_shot_gen_prompt}
\begin{lstlisting}[
  language=Markdown,
  basicstyle=\ttfamily\scriptsize,
  breaklines=false,
  keepspaces=true,
  columns=fullflexible,
  numbers=left,
  numberstyle=\tiny,
  numbersep=4pt,
  xleftmargin=16pt
]
You are an expert API tester who writes pytest tests
to validate that API implementations follow their
intended human-written requirements. [...]

## Requirement:
<REQUIREMENT_TEXT>

## OpenAPI Specification:
<OAS_JSON>

## Instructions (must follow exactly):
- Write a complete pytest test function [...]
- Use the requests library for HTTP calls
- The test MUST be completely self-contained
- Hardcode all required values
  (API base URL should be <SERVICE_BASE_URL>, ...)

Generate ONLY the Python test code, no explanations.
\end{lstlisting}
\end{listing}

\subsection{Experiment 2: Refinement-Loop Generation}
\label{sec:exp2}

The second experiment evaluates a refinement-based strategy that extends the single-step approach with a feedback loop. Generation begins with the same initial prompt used in Experiment~1. The generated test is then executed against the \gls{SUT} while a \textit{Requests\-Tracer} captures all HTTP request/response exchanges. A judge---a second inference call to the same model---receives the requirement, the \gls{OAS}, the generated test code, the test execution output, and the HTTP trace, and returns one of two verdicts (Listing~\ref{lst:judge_prompt}): \texttt{FINISH} if the test is adequate, or \texttt{REFINEMENT} if another iteration is needed. In the latter case, the generator receives the previous test code, the HTTP trace, and the test execution output and produces a revised test, which is then re-evaluated. This process repeats until the judge issues \texttt{FINISH} or the maximum number of rounds is reached (three rounds in our experiments). Note that the judge solely issues a binary continue/stop decision and is part of the generation approach, not the external evaluation; using the same model is thus a deliberate choice to evaluate the model performance.

\begin{listing}[!t]
\caption{Shortened judge prompt template. The judge receives test execution results and decides whether refinement is needed.}
\label{lst:judge_prompt}

\begin{lstlisting}[
  language=Markdown,
  basicstyle=\ttfamily\scriptsize,
  breaklines=false,
  keepspaces=true,
  columns=fullflexible,
  numbers=left,
  numberstyle=\tiny,
  numbersep=4pt,
  xleftmargin=16pt
]
You are a test quality reviewer. [...]

## Requirement Being Tested:
<REQUIREMENT_TEXT>

## Test Code:
<GENERATED_TEST_CODE>

## Test Execution Output:
<STDOUT_STDERR>

## HTTP Trace (actual requests/responses):
<TRACE_JSON>

## Your Task:
1. Compare the test case against the requirement
2. Check if the Test Execution Output and HTTP Trace
   can enhance the test's effectiveness
3. Identify if failures are due to test defects
   OR correct detection of faulty service behaviour
[...]

## Response Format (EXACTLY one of):
REFINEMENT: <short description of what needs changing>
FINISH: <short description why test is correct>
\end{lstlisting}
\end{listing}

\paragraph{Results.}
Figure~\ref{fig:zs_vs_refinement_vague} shows that refinement generally increases mutation scores over single-step generation. The only notable exception is Sonnet 4.5 on precise requirements, where the single-step score already reaches 92\% and refinement yields no further improvement. In general, the gains from refinement are substantially larger for vague requirements than for precise ones. For example, GPT-5 Nano improves by 29 percentage points on vague requirements, reaching 80\%, and GPT-5 Mini improves by 24 percentage points, reaching 75\%.

A consistent pattern across all models is that \(MS_\phi^{m}\) remains below \(MS_\phi^{\text{valid}}\). This indicates that generating tests against a mutated \gls{SUT} systematically reduces test effectiveness. The gap is larger for vague requirements than for precise ones, suggesting that underspecified requirements make refinement-based generation more susceptible to adapting to faulty behaviour. Notably, for four models on precise requirements---Sonnet 4.5, GPT-5 Nano, Mistral Medium, and Llama 3.3 70B---the single-step score exceeds \(MS_\phi^{m}\). In these cases, single-step generation without \gls{SUT} interaction is more effective than refinement performed against a faulty implementation. A plausible explanation is that the \gls{LLM} adapts its assertions to match the faulty behaviour and thereby treats the defect as correct.

\subsection{Costs}
\begin{figure}[h]
    \centering
    \includegraphics[width=\columnwidth]{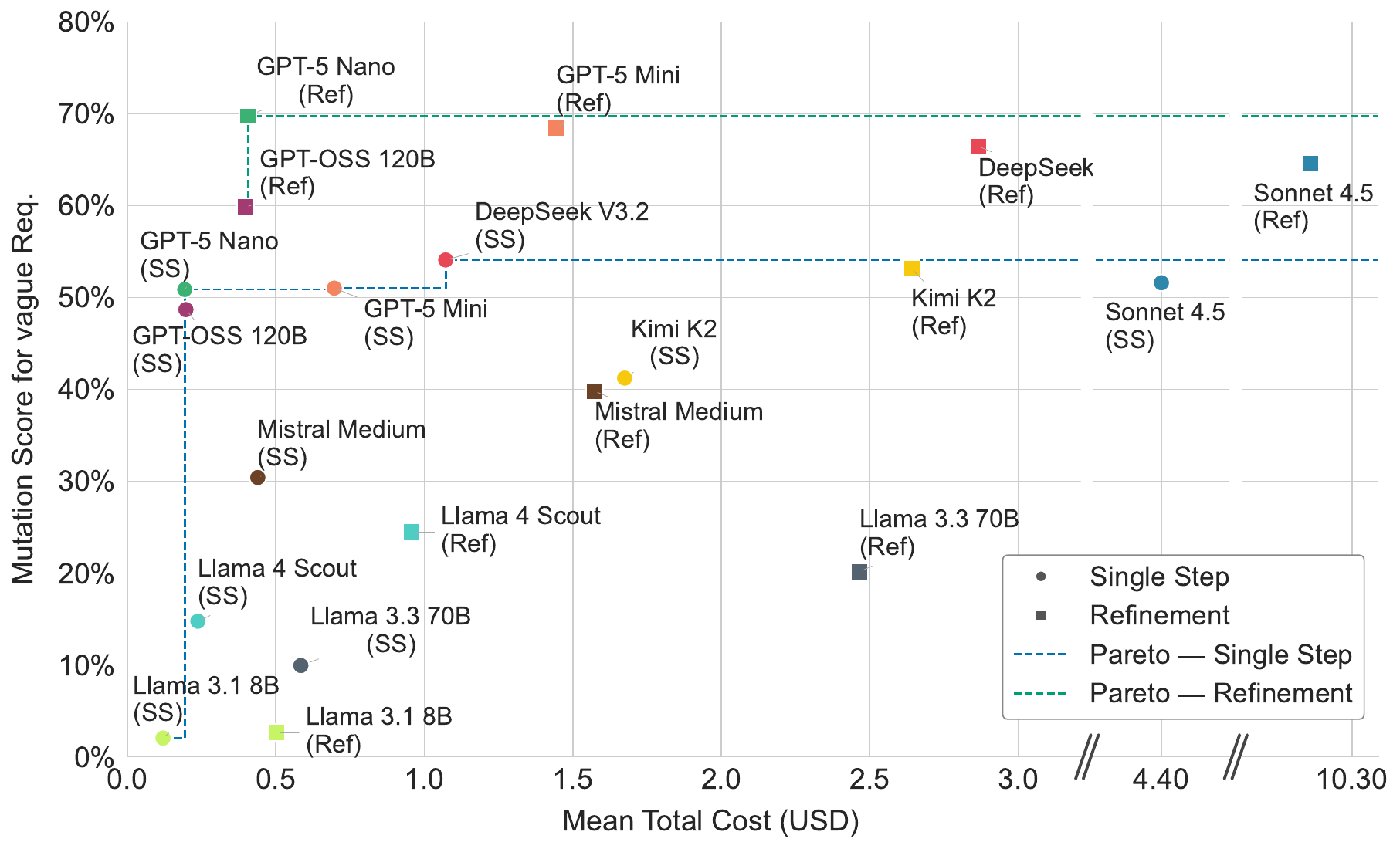}
    \caption{Mean total generation cost (USD) per benchmark run versus mutation score for all models on \emph{vague} requirements under single-step (circles) and refinement (squares, scores and cost averaged over valid and mutated \gls{SUT} conditions).}
    \label{fig:cost_vs_msscore}
\end{figure}

Figure~\ref{fig:cost_vs_msscore} shows that all refinement configurations achieve higher mutation scores than their single-step counterparts, but at increased cost. Across models, refinement increases total cost by roughly 2--4$\times$. However, the gains differ substantially by model.

The strongest cost-effectiveness results are achieved by GPT-5 Nano, GPT-OSS 120B, and GPT-5 Mini. In the refinement setting, GPT-5 Nano reaches a 70\% mutation score at \$0.41 per run, surpassing Sonnet 4.5, which reaches 65\% at \$10.13 per run. This corresponds to a 25$\times$ cost reduction while also achieving higher effectiveness. Similarly, GPT-5 Mini reaches 68\% at \$1.44 per run, making it about 7$\times$ cheaper than Sonnet 4.5 while also outperforming it. More broadly, GPT-5 Nano (+19 pp) and GPT-5 Mini (+17 pp) benefit most from refinement on vague requirements, whereas models such as DeepSeek V3.2 (+12 pp) and Llama 4 Scout (+10 pp) improve more moderately. Sonnet 4.5 improves by 13 percentage points, but at a 2.3$\times$ cost increase (from \$4.42 to \$10.13 per run).

\section{Discussion}

RESTestBench was designed to evaluate the ability of REST API test generation tools to perform \textit{requirements-based test generation}. Current approaches often start from the \gls{OAS} and \gls{SUT} implementation to generate tests, attempting to implicitly or explicitly reverse-engineer requirements from these artifacts as part of the test generation process (Section~\ref{sec:generationApproaches}). Requirements-based tests should correctly differentiate between valid and faulty systems, which is only possible if the requirements are not assumed to correspond to the \gls{SUT}'s behaviour --- in contrast to regression testing which assumes the current implementation to be correct. By having explicit requirements our framework allows us to measure the effect of using information about the \gls{SUT}'s behaviour on requirements-based test generation. Accordingly, we expect \(MS_\phi^m\) to be zero for pure regression test generation approaches, as they would classify a faulty system as correct.

In general, a test suite is considered good if it achieves a mutation score of 90\% or more~\cite{ammannIntroductionSoftwareTesting2017}. In our results (Figure~\ref{fig:zs_vs_refinement_vague}), only Sonnet~4.5, GPT-5~Nano, and GPT-5~Mini exceed this threshold on precise requirements. Since such detailed requirements are rare in real-world settings or legacy code bases, the mutation score on vague requirements provides a better indicator of real-world usefulness. Nevertheless, no model achieved more than 90\% on vague requirements, suggesting that revealing real bugs remains challenging for state-of-the-art models.

Despite its strong single-step performance, Sonnet~4.5's high cost often outweighs its benefits. Smaller models such as GPT-5~Nano or GPT-OSS~120B, when used in a refinement loop, achieve comparable or superior performance at minimal cost. This highlights that precise, human-defined requirements can substantially improve test quality, and that refinement loops enable smaller models to compensate for weaker initial reasoning.

We observe that, for precise requirements, interacting with the \gls{SUT} only slightly increases---or in some cases even decreases---the mutation score, as observed for Sonnet 4.5. One possible explanation is that exposing the actual behaviour of the \gls{SUT} encourages models to validate aspects beyond the specified requirement, producing overly strong oracles that exceed the intended scope and consequently increase false positives, thereby reducing correctness.

A closer examination of the false positive outputs reveals a qualitative gap in how models fail. Weak models such as the three Llama variants exhibit \textit{construction-level} failures: asserting wrong HTTP status codes, hallucinating non-existent functions or API endpoints, and leaving hardcoded placeholder values. In contrast, Sonnet~4.5 produces virtually no code errors but is more prone to \textit{specification-interpretation} mismatches---assuming standard API response fields that this implementation omits, or expecting conventional status codes (e.g.\ $422$) where the service returns $400$. This underscores that the quality and detail of the requirement specification plays a critical role: weak models fail to construct valid tests regardless of specification quality, while strong models fail when the requirement leaves room for interpretation, causing them to fall back on general REST conventions rather than implementation-specific behaviour.

Comparing \(MS_\phi^{\text{valid}}\) with \(MS_\phi^m\) confirms the observation of~\cite{konstantinouLLMsGenerateTest2024} that \gls{LLM} test oracle accuracy considerably drops in the presence of buggy code, suggesting that predictions are biased towards the actual implementation rather than the desired one. Taking requirement detail into account, we observe that this drop is also larger on vague requirements. Moreover, tests generated on mutated code can negate the benefit of a refinement approach to such an extent that a non-refinement approach becomes more effective, as seen with Sonnet~4.5, GPT-5~Nano, Mistral~Medium, and Llama~3.3~70B. This underscores that, for these models, incorporating actual \gls{SUT} behaviour is unnecessary when requirement detail is high.

While RESTestBench provides a manually validated set of requirements, most tools surveyed in Section~\ref{sec:generationApproaches} derive requirements automatically using \glspl{LLM} without validation, suggesting that test correctness may further decrease when relying on unverified requirements.

\section{Threats to Validity}

\textbf{Internal Validity.} Several factors in our experimental setup could confound the observed results. First, prompt design can substantially influence model behaviour. Second, OpenAPI specification mismatches may exist, which can affect reported mutation scores, especially for more capable models.

\textbf{Construct Validity.} Our measures rely on proxies for true software correctness and test effectiveness. The mutation operators are human-defined decisions about what constitute meaningful requirement violations, and the mutation-based metrics approximate the effectiveness of requirements-based test generation and oracle quality. Furthermore, we cannot algorithmically determine whether a failing requirements-based test reflects an invalid test or a latent defect, due to the undecidability of non-trivial program properties~\cite{riceClassesRecursivelyEnumerable1953}. To mitigate this, we created a golden test set covering obvious correct behaviours.

\textbf{External Validity.} Our benchmark includes only three services, which may not represent the full diversity and complexity of industrial software projects. Additionally, all generated tests were implemented in Python; results may differ under other programming languages, frameworks, or execution environments.

\textbf{Conclusion Validity.} Each experiment was executed only three times, limiting the statistical robustness of our findings.

\section{Conclusion}
Since the emergence of \glspl{LLM}, several studies have investigated their effectiveness in \gls{REST} API testing. Unfortunately, their effectiveness has so far been evaluated primarily using code-based coverage and fault-detection metrics, which provide only limited insight into whether generated tests actually validate the intended functional behaviour of a service. With \textit{RESTestBench}, we provide a means to close this gap by enabling the systematic and reproducible evaluation of generated tests against human-validated behavioural expectations and manually defined, requirement-aligned mutations, thereby building on and extending the \gls{PBMT} framework introduced by Bartocci et al.~\cite{bartocciPropertyBasedMutationTesting2023}. To illustrate this potential and to revisit assumptions implicit in earlier tool-oriented approaches, we conducted two controlled experiments showing that (1) more precise requirement descriptions lead to substantially more effective tests, (2) refinement loops can improve effectiveness, especially in underspecified settings, (3) interaction with a faulty implementation may negatively influence the resulting tests by causing generation to adapt to incorrect behaviour, and (4) while small models may fail to reach mutated code sections, larger models can over-engineer their test oracles, verifying beyond the requirement and reducing overall scores. These findings motivate future work to use the benchmark for comparing new test-generation strategies, investigating the influence of OpenAPI structure and quality, analyzing effectiveness--cost trade-offs across \gls{LLM}-based approach and a lot more. In addition, we aim to extend the benchmark with other services and add microservice-oriented systems, such as \textit{Train Ticket} in order to evaluate requirement-based test generation in more realistic multi-service settings.

\begin{acks}
This work was supported by and done within the scope of the ITEA4 GENIUS project, which was nationally funded by FFG with grant 921454. 
\end{acks}

\bibliographystyle{ACM-Reference-Format}
\bibliography{bibliography}

\end{document}

%% file: tables/tool_overview.tex
\begin{table}[tb]
\centering
\caption{Overview of LLM-based REST API test generation tools. 
Ref. = Uses refinement loop, CC = Code Coverage, MC = Mutation Coverage.}
\label{tab:tool_overview}
\footnotesize
\setlength{\tabcolsep}{3pt}
\begin{tabular}{p{2.2cm} p{2.6cm} c p{2.5cm}}
\toprule
\textbf{Tool} & \textbf{Oracle Source} & \textbf{Ref.} & \textbf{Evaluation} \\
\midrule
APITestGenie~\cite{pereiraAPITestGenieAutomatedAPI2024a}
& Human-defined NL req. & \xmark & human \\

LogiAgent~\cite{zhangLogiAgentAutomatedLogical2025a}
& From OAS (LLM) & \cmark & CC + \#5xx + human \\

RestTSLLM~\cite{barradasCombiningTSLLLM2025}
& From OAS (LLM) & \xmark & CC + MC \\

SAINT~\cite{panSAINTServicelevelIntegration2025}
& From code (LLM) & \cmark & CC + \#5xx + human \\

RESTifAI~\cite{koglerRESTifAILLMBasedWorkflow2025}
& From OAS (LLM) & \cmark & CC + \#5xx + human \\
\bottomrule
\end{tabular}
\end{table}

%% file: tables/list_of_services.tex
\begin{table}[h]
\centering
\small
\caption{Services used in RESTestBench. Stars/Forks are taken from the exact GitHub repositories (accessed on 2026--01--29). Endpoint counts and CLOC (counted lines of code) are computed from the corresponding benchmark snapshot.}
\begin{tabular}{l c r r r}\toprule
\textbf{Name} & \shortstack{\textbf{GET/POST/PUT}\\\textbf{PATCH/DELETE}} & \textbf{Stars} & \textbf{Forks} & \textbf{CLOC} \\
\midrule
FastAPI \cite{FastapiFullstackfastapitemplate2026} &
6/10/1/3/3 & 41.2k & 8k & 8,072 \\
TodoApp \cite{fowlerDavidfowlTodoApp2026} &
2/3/1/0/1 & 3.1k & 462 & 1,609 \\
RealWorld \cite{jakobLujakobNestjsrealworldexampleapp2026} &
7/6/2/0/4 & 3.3k & 701 & 965 \\
\bottomrule
\end{tabular}
\label{tab:services}
\end{table}

%% file: tables/overview_nr_req_mutations.tex
\begin{table}[h]
\centering
\small
\caption{Requirements and mutations per service used in the benchmark.}
\begin{tabular}{l c c c}
\toprule
\textbf{Service} & \textbf{\#req.} & \textbf{\#mut.} & \textbf{avg. mut/req} \\
\midrule
FastAPI \cite{FastapiFullstackfastapitemplate2026} & 63 & 141 & 2.24 \\
TodoApp \cite{fowlerDavidfowlTodoApp2026} & 14 & 22 & 1.57 \\
RealWorld \cite{jakobLujakobNestjsrealworldexampleapp2026} & 29 & 65 & 2.24 \\
\midrule
\textbf{Total} & \textbf{106} & \textbf{228} & \textbf{2.15} \\
\bottomrule
\end{tabular}
\label{tab:overview_requirements}
\end{table}

%% file: tables/list_of_models.tex
\begin{table}[h]
\centering
\small
\setlength{\tabcolsep}{3.5pt} 
\caption{LLM models used in the benchmark with licensing type, parameter count, and cost per million tokens (Azure AI Foundry pricing).}
\begin{tabular}{c l l r r}
\toprule
 & \textbf{Model} & \textbf{Company} & \textbf{Size} & \textbf{\$/1M (In/Out)} \\
\midrule
\multirow{6}{*}{\rotatebox[origin=c]{90}{\scriptsize\textbf{Open Source}}}
 & Llama 3.1 8B        & Meta        & 8B            & 0.15 / 0.15 \\
 & Llama 3.3 70B       & Meta        & 70B           & 0.71 / 0.71 \\
 & Llama 4 Scout       & Meta        & 109B (17B)    & 0.25 / 1.00 \\
 & GPT-OSS 120B        & OpenAI      & 117B (5.1B)   & 0.15 / 0.60 \\
 & DeepSeek V3.2       & DeepSeek    & 671B (37B)    & 0.58 / 1.68 \\
 & Kimi K2 Thinking    & Moonshot AI & 1T (32B)      & 0.60 / 2.50 \\
\midrule
\multirow{4}{*}{\rotatebox[origin=c]{90}{\scriptsize\textbf{Closed Source}}}
 & GPT-5 Nano          & OpenAI      & --            & 0.05 / 0.40 \\
 & GPT-5 Mini          & OpenAI      & --            & 0.25 / 2.00 \\
 & Mistral Medium      & Mistral AI  & --            & 0.40 / 2.00 \\
 & Claude Sonnet 4.5   & Anthropic   & --            & 3.00 / 15.00 \\
\bottomrule
\end{tabular}
\label{tab:llm_models}
\end{table}

%% file: bibliography.bib
@inproceedings{alonsoAGORAAutomatedGeneration2023,
  title = {{{AGORA}}: {{Automated Generation}} of {{Test Oracles}} for {{REST APIs}}},
  shorttitle = {{{AGORA}}},
  booktitle = {Proceedings of the 32nd {{ACM SIGSOFT International Symposium}} on {{Software Testing}} and {{Analysis}}},
  author = {Alonso, Juan C. and Segura, Sergio and {Ruiz-Cort{\'e}s}, Antonio},
  year = 2023,
  month = jul,
  pages = {1018--1030},
  publisher = {ACM},
  address = {Seattle WA USA},
  doi = {10.1145/3597926.3598114},
  urldate = {2026-02-03},
  isbn = {979-8-4007-0221-1},
  langid = {english}
}

@book{ammannIntroductionSoftwareTesting2017,
  title = {Introduction to Software Testing},
  author = {Ammann, Paul and Offutt, Jeff},
  year = 2017,
  edition = {2nd ed},
  publisher = {Cambridge university press},
  address = {Cambridge},
  isbn = {978-1-107-17201-2},
  langid = {english},
  lccn = {004.24}
}

@misc{andrewsOpenAPISpecification2026,
  title = {{{OpenAPI Specification}}},
  author = {Andrews, Henry},
  year = 2026,
  month = feb
}

@article{arcuriAutomatedBlackWhiteBox2021,
  title = {Automated {{Black-}} and {{White-Box Testing}} of {{RESTful APIs With EvoMaster}}},
  author = {Arcuri, Andrea},
  year = 2021,
  month = may,
  journal = {IEEE Software},
  volume = {38},
  number = {3},
  pages = {72--78},
  issn = {0740-7459, 1937-4194},
  doi = {10.1109/MS.2020.3013820},
  urldate = {2025-12-18},
  copyright = {https://ieeexplore.ieee.org/Xplorehelp/downloads/license-information/IEEE.html}
}

@inproceedings{aroraGeneratingTestScenarios2024,
  title = {Generating {{Test Scenarios}} from {{NL Requirements}} Using {{Retrieval-Augmented LLMs}}: {{An Industrial Study}}},
  shorttitle = {Generating {{Test Scenarios}} from {{NL Requirements}} Using {{Retrieval-Augmented LLMs}}},
  booktitle = {2024 {{IEEE}} 32nd {{International Requirements Engineering Conference}} ({{RE}})},
  author = {Arora, Chetan and Herda, Tomas and Homm, Verena},
  year = 2024,
  month = jun,
  eprint = {2404.12772},
  primaryclass = {cs},
  pages = {240--251},
  doi = {10.1109/RE59067.2024.00031},
  urldate = {2026-02-26},
  archiveprefix = {arXiv}
}

@inproceedings{atlidakisRESTlerStatefulREST2019,
  title = {{{RESTler}}: {{Stateful REST API Fuzzing}}},
  shorttitle = {{{RESTler}}},
  booktitle = {2019 {{IEEE}}/{{ACM}} 41st {{International Conference}} on {{Software Engineering}} ({{ICSE}})},
  author = {Atlidakis, Vaggelis and Godefroid, Patrice and Polishchuk, Marina},
  year = 2019,
  month = may,
  pages = {748--758},
  publisher = {IEEE},
  address = {Montreal, QC, Canada},
  doi = {10.1109/ICSE.2019.00083},
  urldate = {2026-02-03},
  copyright = {https://ieeexplore.ieee.org/Xplorehelp/downloads/license-information/IEEE.html},
  isbn = {978-1-7281-0869-8}
}

@misc{augustoLargeLanguageModels2025,
  title = {Large {{Language Models}} for {{Software Testing}}: {{A Research Roadmap}}},
  shorttitle = {Large {{Language Models}} for {{Software Testing}}},
  author = {Augusto, Cristian and Bertolino, Antonia and De Angelis, Guglielmo and Lonetti, Francesca and Mor{\'a}n, Jes{\'u}s},
  year = 2025,
  publisher = {arXiv},
  doi = {10.48550/ARXIV.2509.25043},
  urldate = {2025-12-18},
  copyright = {Creative Commons Attribution 4.0 International}
}

@misc{barradasCombiningTSLLLM2025,
  title = {Combining {{TSL}} and {{LLM}} to {{Automate REST API Testing}}: {{A Comparative Study}}},
  shorttitle = {Combining {{TSL}} and {{LLM}} to {{Automate REST API Testing}}},
  author = {Barradas, Thiago and Paes, Aline and Neves, V{\^a}nia de Oliveira},
  year = 2025,
  publisher = {arXiv},
  doi = {10.48550/ARXIV.2509.05540},
  urldate = {2026-01-13},
  copyright = {Creative Commons Attribution 4.0 International}
}

@article{barrOracleProblemSoftware2015,
  title = {The {{Oracle Problem}} in {{Software Testing}}: {{A Survey}}},
  shorttitle = {The {{Oracle Problem}} in {{Software Testing}}},
  author = {Barr, Earl T. and Harman, Mark and McMinn, Phil and Shahbaz, Muzammil and Yoo, Shin},
  year = 2015,
  month = may,
  journal = {IEEE Transactions on Software Engineering},
  volume = {41},
  number = {5},
  pages = {507--525},
  issn = {0098-5589, 1939-3520},
  doi = {10.1109/TSE.2014.2372785},
  urldate = {2025-12-15},
  copyright = {https://creativecommons.org/licenses/by/3.0/legalcode}
}

@inproceedings{bartocciPropertyBasedMutationTesting2023,
  title = {Property-{{Based Mutation Testing}}},
  booktitle = {2023 {{IEEE Conference}} on {{Software Testing}}, {{Verification}} and {{Validation}} ({{ICST}})},
  author = {Bartocci, Ezio and Mariani, Leonardo and Ni{\v c}kovi{\'c}, Dejan and Yadav, Drishti},
  year = 2023,
  month = apr,
  pages = {222--233},
  publisher = {IEEE},
  address = {Dublin, Ireland},
  doi = {10.1109/ICST57152.2023.00029},
  urldate = {2026-02-10},
  copyright = {https://doi.org/10.15223/policy-029},
  isbn = {978-1-6654-5666-1}
}

@book{cockburnWritingEffectiveUse2012,
  title = {Writing Effective Use Cases},
  author = {Cockburn, Alistair},
  year = 2012,
  series = {The {{Agile}} Software Development Series},
  edition = {24. print},
  publisher = {Addison-Wesley},
  address = {Boston},
  isbn = {978-0-201-70225-5},
  langid = {english}
}

@inproceedings{colesPITPracticalMutation2016,
  title = {{{PIT}}: A Practical Mutation Testing Tool for {{Java}} (Demo)},
  shorttitle = {{{PIT}}},
  booktitle = {Proceedings of the 25th {{International Symposium}} on {{Software Testing}} and {{Analysis}}},
  author = {Coles, Henry and Laurent, Thomas and Henard, Christopher and Papadakis, Mike and Ventresque, Anthony},
  year = 2016,
  month = jul,
  pages = {449--452},
  publisher = {ACM},
  address = {Saarbr\"ucken Germany},
  doi = {10.1145/2931037.2948707},
  urldate = {2026-02-10},
  isbn = {978-1-4503-4390-9},
  langid = {english}
}

@inproceedings{decropPublicBenchmarkREST2025,
  title = {A {{Public Benchmark}} of {{REST APIs}}},
  booktitle = {2025 {{IEEE}}/{{ACM}} 22nd {{International Conference}} on {{Mining Software Repositories}} ({{MSR}})},
  author = {Decrop, Alix and Eraso, Sara and Devroey, Xavier and Perrouin, Gilles},
  year = 2025,
  month = apr,
  pages = {421--433},
  publisher = {IEEE},
  address = {Ottawa, ON, Canada},
  doi = {10.1109/MSR66628.2025.00072},
  urldate = {2025-12-15},
  copyright = {https://doi.org/10.15223/policy-029},
  isbn = {979-8-3315-0183-9}
}

@incollection{demeyerFormalVerificationDeveloper2020,
  title = {Formal {{Verification}} of {{Developer Tests}}: {{A Research Agenda Inspired}} by {{Mutation Testing}}},
  shorttitle = {Formal {{Verification}} of {{Developer Tests}}},
  booktitle = {Leveraging {{Applications}} of {{Formal Methods}}, {{Verification}} and {{Validation}}: {{Engineering Principles}}},
  author = {Demeyer, Serge and Parsai, Ali and Vercammen, Sten and Van Bladel, Brent and Abdi, Mehrdad},
  editor = {Margaria, Tiziana and Steffen, Bernhard},
  year = 2020,
  volume = {12477},
  pages = {9--24},
  publisher = {Springer International Publishing},
  address = {Cham},
  doi = {10.1007/978-3-030-61470-6_2},
  urldate = {2026-02-23},
  isbn = {978-3-030-61469-0 978-3-030-61470-6},
  langid = {english}
}

@misc{FastapiFullstackfastapitemplate2026,
  title = {Fastapi/Full-Stack-Fastapi-Template},
  year = 2026,
  month = jan,
  urldate = {2026-01-29},
  copyright = {MIT},
  howpublished = {FastAPI}
}

@misc{fowlerDavidfowlTodoApp2026,
  title = {Davidfowl/{{TodoApp}}},
  author = {Fowler, David},
  year = 2026,
  month = jan,
  urldate = {2026-01-29},
  copyright = {MIT}
}

@article{golmohammadiTestingRESTfulAPIs2024,
  title = {Testing {{RESTful APIs}}: {{A Survey}}},
  shorttitle = {Testing {{RESTful APIs}}},
  author = {Golmohammadi, Amid and Zhang, Man and Arcuri, Andrea},
  year = 2024,
  month = jan,
  journal = {ACM Transactions on Software Engineering and Methodology},
  volume = {33},
  number = {1},
  pages = {1--41},
  publisher = {Association for Computing Machinery (ACM)},
  issn = {1049-331X, 1557-7392},
  doi = {10.1145/3617175},
  urldate = {2025-07-08},
  copyright = {https://creativecommons.org/licenses/by/4.0/},
  langid = {english}
}

@article{golmohammadiTestingRESTfulAPIs2024a,
  title = {Testing {{RESTful APIs}}: {{A Survey}}},
  shorttitle = {Testing {{RESTful APIs}}},
  author = {Golmohammadi, Amid and Zhang, Man and Arcuri, Andrea},
  year = 2024,
  month = jan,
  journal = {ACM Transactions on Software Engineering and Methodology},
  volume = {33},
  number = {1},
  pages = {1--41},
  issn = {1049-331X, 1557-7392},
  doi = {10.1145/3617175},
  urldate = {2025-12-15},
  langid = {english}
}

@inproceedings{groceExtensibleRegularexpressionbasedTool2018,
  title = {An Extensible, Regular-Expression-Based Tool for Multi-Language Mutant Generation},
  booktitle = {Proceedings of the 40th {{International Conference}} on {{Software Engineering}}: {{Companion Proceeedings}}},
  author = {Groce, Alex and Holmes, Josie and Marinov, Darko and Shi, August and Zhang, Lingming},
  year = 2018,
  month = may,
  pages = {25--28},
  publisher = {ACM},
  address = {Gothenburg Sweden},
  doi = {10.1145/3183440.3183485},
  urldate = {2026-02-10},
  isbn = {978-1-4503-5663-3},
  langid = {english}
}

@misc{hossainBriefSurveyOraclebased2022,
  title = {A {{Brief Survey}} on {{Oracle-based Test Adequacy Metrics}}},
  author = {Hossain, Soneya Binta and Dwyer, Matthew B.},
  year = 2022,
  publisher = {arXiv},
  doi = {10.48550/ARXIV.2212.06118},
  urldate = {2025-12-15},
  copyright = {Creative Commons Attribution 4.0 International}
}

@misc{huangChallengesFuzzingTechniques2024,
  title = {On the {{Challenges}} of {{Fuzzing Techniques}} via {{Large Language Models}}},
  author = {Huang, Linghan and Zhao, Peizhou and Chen, Huaming and Ma, Lei},
  year = 2024,
  publisher = {arXiv},
  doi = {10.48550/ARXIV.2402.00350},
  urldate = {2025-12-18},
  copyright = {arXiv.org perpetual, non-exclusive license}
}

@inproceedings{inozemtsevaCoverageNotStrongly2014,
  title = {Coverage Is Not Strongly Correlated with Test Suite Effectiveness},
  booktitle = {Proceedings of the 36th {{International Conference}} on {{Software Engineering}}},
  author = {Inozemtseva, Laura and Holmes, Reid},
  year = 2014,
  month = may,
  pages = {435--445},
  publisher = {ACM},
  address = {Hyderabad India},
  doi = {10.1145/2568225.2568271},
  urldate = {2026-02-09},
  isbn = {978-1-4503-2756-5},
  langid = {english}
}

@book{jacobsonObjectorientedSoftwareEngineering2005,
  title = {Object-Oriented Software Engineering: A Use Case Driven Approach},
  shorttitle = {Object-Oriented Software Engineering},
  author = {Jacobson, Ivar},
  year = 2005,
  publisher = {Addison-Wesley Pub},
  address = {Reading, Mass.},
  isbn = {978-0-201-40347-3},
  langid = {english}
}

@misc{jakobLujakobNestjsrealworldexampleapp2026,
  title = {Lujakob/Nestjs-Realworld-Example-App},
  author = {Jakob, Lukas},
  year = 2026,
  month = jan,
  urldate = {2026-01-29}
}

@misc{koglerRESTifAILLMBasedWorkflow2025,
  title = {{{RESTifAI}}: {{LLM-Based Workflow}} for {{Reusable REST API Testing}}},
  shorttitle = {{{RESTifAI}}},
  author = {Kogler, Leon and Ehrhart, Maximilian and Dornauer, Benedikt and Enoiu, Eduard Paul},
  year = 2025,
  publisher = {arXiv},
  doi = {10.48550/ARXIV.2512.08706},
  urldate = {2025-12-15},
  copyright = {Creative Commons Attribution 4.0 International}
}

@misc{konstantinouLLMsGenerateTest2024,
  title = {Do {{LLMs}} Generate Test Oracles That Capture the Actual or the Expected Program Behaviour?},
  author = {Konstantinou, Michael and Degiovanni, Renzo and Papadakis, Mike},
  year = 2024,
  publisher = {arXiv},
  doi = {10.48550/ARXIV.2410.21136},
  urldate = {2025-12-15},
  copyright = {Creative Commons Attribution 4.0 International}
}

@article{liTestOracleStrategies2017,
  title = {Test {{Oracle Strategies}} for {{Model-Based Testing}}},
  author = {Li, Nan and Offutt, Jeff},
  year = 2017,
  month = apr,
  journal = {IEEE Transactions on Software Engineering},
  volume = {43},
  number = {4},
  pages = {372--395},
  issn = {0098-5589, 1939-3520},
  doi = {10.1109/TSE.2016.2597136},
  urldate = {2026-02-09},
  copyright = {https://ieeexplore.ieee.org/Xplorehelp/downloads/license-information/IEEE.html}
}

@inproceedings{martin-lopezRESTestAutomatedBlackbox2021,
  title = {{{RESTest}}: Automated Black-Box Testing of {{RESTful}} Web {{APIs}}},
  shorttitle = {{{RESTest}}},
  booktitle = {Proceedings of the 30th {{ACM SIGSOFT International Symposium}} on {{Software Testing}} and {{Analysis}}},
  author = {{Martin-Lopez}, Alberto and Segura, Sergio and {Ruiz-Cort{\'e}s}, Antonio},
  year = 2021,
  month = jul,
  pages = {682--685},
  publisher = {ACM},
  address = {Virtual Denmark},
  doi = {10.1145/3460319.3469082},
  urldate = {2026-02-03},
  isbn = {978-1-4503-8459-9},
  langid = {english}
}

@article{mohammedmudassirRoleAPIsModern2024,
  title = {The Role of {{APIs}} in Modern Software Development},
  author = {{Mohammed Mudassir} and {Mohammed Mushtaq}},
  year = 2024,
  month = oct,
  journal = {World Journal of Advanced Engineering Technology and Sciences},
  volume = {13},
  number = {1},
  pages = {1045--1047},
  issn = {25828266},
  doi = {10.30574/wjaets.2024.13.1.0515},
  urldate = {2025-12-17}
}

@incollection{MutationTestingAdvances2019,
  title = {Mutation {{Testing Advances}}: {{An Analysis}} and {{Survey}}},
  shorttitle = {Mutation {{Testing Advances}}},
  booktitle = {Advances in {{Computers}}},
  author = {Papadakis, Mike and Kintis, Marinos and Zhang, Jie and Jia, Yue and Le Traon, Yves and Harman, Mark},
  year = 2019,
  volume = {112},
  pages = {275--378},
  publisher = {Elsevier},
  doi = {10.1016/bs.adcom.2018.03.015},
  urldate = {2026-02-10},
  copyright = {https://www.elsevier.com/tdm/userlicense/1.0/},
  isbn = {978-0-12-815121-1},
  langid = {english}
}

@misc{panSAINTServicelevelIntegration2025,
  title = {{{SAINT}}: {{Service-level Integration Test Generation}} with {{Program Analysis}} and {{LLM-based Agents}}},
  shorttitle = {{{SAINT}}},
  author = {Pan, Rangeet and Pavuluri, Raju and Huang, Ruikai and Krishna, Rahul and Stennett, Tyler and Orso, Alessandro and SInha, Saurabh},
  year = 2025,
  publisher = {arXiv},
  doi = {10.48550/ARXIV.2511.13305},
  urldate = {2025-12-11},
  copyright = {Creative Commons Attribution Non Commercial No Derivatives 4.0 International}
}

@inproceedings{papadakisAreMutationScores2018a,
  title = {Are Mutation Scores Correlated with Real Fault Detection?: A Large Scale Empirical Study on the Relationship between Mutants and Real Faults},
  shorttitle = {Are Mutation Scores Correlated with Real Fault Detection?},
  booktitle = {Proceedings of the 40th {{International Conference}} on {{Software Engineering}}},
  author = {Papadakis, Mike and Shin, Donghwan and Yoo, Shin and Bae, Doo-Hwan},
  year = 2018,
  month = may,
  pages = {537--548},
  publisher = {ACM},
  address = {Gothenburg Sweden},
  doi = {10.1145/3180155.3180183},
  urldate = {2026-02-10},
  isbn = {978-1-4503-5638-1},
  langid = {english}
}

@misc{pereiraAPITestGenieAutomatedAPI2024a,
  title = {{{APITestGenie}}: {{Automated API Test Generation}} through {{Generative AI}}},
  shorttitle = {{{APITestGenie}}},
  author = {Pereira, Andr{\'e} and Lima, Bruno and Faria, Jo{\~a}o Pascoal},
  year = 2024,
  publisher = {arXiv},
  doi = {10.48550/ARXIV.2409.03838},
  urldate = {2025-12-15},
  copyright = {arXiv.org perpetual, non-exclusive license}
}

@techreport{postmanStateAPIReport,
  title = {State of the {{API Report}}},
  author = {Postman}
}

@misc{RealworldappsRealworldMother,
  title = {Realworld-Apps/Realworld: "{{The}} Mother of All Demo Apps" --- {{Exemplary}} Fullstack {{Medium}}.Com Clone Powered by {{React}}, {{Angular}}, {{Node}}, {{Django}}, and Many More},
  urldate = {2026-02-09},
  howpublished = {https://github.com/realworld-apps/realworld?tab=readme-ov-file}
}

@article{riceClassesRecursivelyEnumerable1953,
  title = {Classes of Recursively Enumerable Sets and Their Decision Problems},
  author = {Rice, H. G.},
  year = 1953,
  journal = {Transactions of the American Mathematical Society},
  volume = {74},
  number = {2},
  pages = {358--366},
  issn = {0002-9947, 1088-6850},
  doi = {10.1090/S0002-9947-1953-0053041-6},
  urldate = {2026-03-02},
  langid = {english}
}

@article{sanchezMutationTestingPractice2024,
  title = {Mutation {{Testing}} in {{Practice}}: {{Insights From Open-Source Software Developers}}},
  shorttitle = {Mutation {{Testing}} in {{Practice}}},
  author = {S{\'a}nchez, Ana B. and Parejo, Jos{\'e} A. and Segura, Sergio and Dur{\'a}n, Amador and Papadakis, Mike},
  year = 2024,
  month = may,
  journal = {IEEE Transactions on Software Engineering},
  volume = {50},
  number = {5},
  pages = {1130--1143},
  issn = {0098-5589, 1939-3520, 2326-3881},
  doi = {10.1109/TSE.2024.3377378},
  urldate = {2026-02-10},
  copyright = {https://creativecommons.org/licenses/by/4.0/legalcode}
}

@inproceedings{vanlamsweerdeGoalorientedRequirementsEngineering2000,
  title = {Goal-Oriented Requirements Engineering: A Guided Tour},
  shorttitle = {Goal-Oriented Requirements Engineering},
  booktitle = {Proceedings {{Fifth IEEE International Symposium}} on {{Requirements Engineering}}},
  author = {Van Lamsweerde, A.},
  year = 2000,
  pages = {249--262},
  publisher = {IEEE Comput. Soc},
  address = {Toronto, Ont., Canada},
  doi = {10.1109/ISRE.2001.948567},
  urldate = {2026-02-26},
  isbn = {978-0-7695-1125-2}
}

@incollection{vanoverbergheStateCoverageSoftware2012,
  title = {State {{Coverage}}: {{Software Validation Metrics}} beyond {{Code Coverage}}},
  shorttitle = {State {{Coverage}}},
  booktitle = {{{SOFSEM}} 2012: {{Theory}} and {{Practice}} of {{Computer Science}}},
  author = {Vanoverberghe, Dries and De Halleux, Jonathan and Tillmann, Nikolai and Piessens, Frank},
  editor = {Bielikov{\'a}, M{\'a}ria and Friedrich, Gerhard and Gottlob, Georg and Katzenbeisser, Stefan and Tur{\'a}n, Gy{\"o}rgy},
  year = 2012,
  volume = {7147},
  pages = {542--553},
  publisher = {Springer Berlin Heidelberg},
  address = {Berlin, Heidelberg},
  doi = {10.1007/978-3-642-27660-6_44},
  urldate = {2026-02-09},
  copyright = {http://www.springer.com/tdm},
  isbn = {978-3-642-27659-0 978-3-642-27660-6},
  langid = {english}
}

@book{wiegersSoftwareRequirements2013,
  title = {Software Requirements},
  author = {Wiegers, Karl E. and Beatty, Joy and Wiegers, Karl E.},
  year = 2013,
  series = {Best Practices},
  edition = {3. ed. [fully updated and expanded]},
  publisher = {Microsoft Press},
  address = {Redmond, Wash},
  isbn = {978-0-7356-7966-5},
  langid = {english}
}

@misc{yangRequirementsBasedTestGeneration2025,
  title = {Requirements-{{Based Test Generation}}: {{A Comprehensive Survey}}},
  shorttitle = {Requirements-{{Based Test Generation}}},
  author = {Yang, Zhenzhen and Huang, Rubing and Cui, Chenhui and Niu, Nan and Towey, Dave},
  year = 2025,
  publisher = {arXiv},
  doi = {10.48550/ARXIV.2505.02015},
  urldate = {2026-01-15},
  copyright = {Creative Commons Attribution 4.0 International}
}

@misc{zhangLogiAgentAutomatedLogical2025a,
  title = {{{LogiAgent}}: {{Automated Logical Testing}} for {{REST Systems}} with {{LLM-Based Multi-Agents}}},
  shorttitle = {{{LogiAgent}}},
  author = {Zhang, Ke and Zhang, Chenxi and Wang, Chong and Zhang, Chi and Wu, YaChen and Xing, Zhenchang and Liu, Yang and Li, Qingshan and Peng, Xin},
  year = 2025,
  publisher = {arXiv},
  doi = {10.48550/ARXIV.2503.15079},
  urldate = {2025-12-15},
  copyright = {Creative Commons Attribution 4.0 International}
}

@article{zowghiErratumInterplayConsistency2004,
  title = {Erratum to ``{{On}} the Interplay between Consistency, Completeness, and Correctness in Requirements Evolution''},
  author = {Zowghi, Didar and Gervasi, Vincenzo},
  year = 2004,
  month = sep,
  journal = {Information and Software Technology},
  volume = {46},
  number = {11},
  pages = {763--779},
  issn = {09505849},
  doi = {10.1016/j.infsof.2004.03.003},
  urldate = {2026-02-26},
  copyright = {https://www.elsevier.com/tdm/userlicense/1.0/},
  langid = {english}
}
